*Chapter 1*

# Hydrogen storage and compression

*Sofoklis S. Makridis*[1,2]

## 1.1 Towards a hydrogen economy

Energy has always been the driving force in the technological and economic development of societies. The consumption of a significant amount of energy is required to provide basic living conditions of developed countries (heating, transportation, lighting, etc.). Today's energy supply has a considerable impact on the environment, since it is fuelled by the burning of fossil fuels. In addition to this, the fossil fuel reserves are decreasing while the demand for energy is rapidly rising. Climate change, the depletion and geographical segregation of fossil fuel resources, health related issues as well as energy poverty constitute the driving forces towards the pursuit of alternative energy sources. In addition, countries with no access to oil reserves are being dependent from other countries for their energy supply, with a strong impact on politics and financial issues. But apart from occasional financial recessions, the long-term need for increasing amounts of energy as countries develop will become a major rate limiting step in the growth of the world economy [1]. The last years there is an on-going research on alternative fuels in order to overcome the fossil energy dependence and to provide a sustainable growth of economies and societies.

In view of the above, countries that release the largest amounts of greenhouse gases to the atmosphere compared to the energy production are expected to minimise $CO_2$ emission and at the same time improve the share of "clean" energy in total energy consumption. The renewable, non-conventional energy sources, such as solar and wind energy, will remain available for infinite period. But due to the inherent nature of renewable energy resources being intermittent, there is a need to store any surplus electrical energy produced in order to be used during high energy

[1]Department of Mechanical Engineering, University of Western Macedonia, GR50132 Kozani, Greece
[2]Department of Environmental and Natural Resources Management, University of Patras, GR30100 Agrinio, Greece

demand periods [2, 3]. The following are the solutions to store surplus energy produced (either in full operation or in an experimental mode) [4–7]:

a. compressed air energy storage
b. batteries
c. flywheel (mechanical inertia) energy storage
d. hydroelectricity (pumped water energy storage)
e. superconducting magnetic energy storage
f. thermal energy storage
g. hydrogen production and then storage or injection into natural gas grid (power to gas)

In this study, only the solutions related to hydrogen are discussed.

As it is derived from the above, hydrogen is an energy carrier and it is related to a process that begins and ends with plain water, known as the "hydrogen fuel cycle." In a more particular manner, water is split into hydrogen and oxygen by the process of electrolysis and then they re-combine, producing electricity and water vapour, using a fuel cell. Furthermore, hydrogen is abundant (e.g. within water) and evenly distributed throughout the world providing security in energy. But even in this case, electricity is needed. Conventional sources of electricity are being used for electrolysis, and as a result the hydrogen carbon footprint remains more or less high. Thus, with renewable energy powered electrolysers (i.e. from solar or aeolian generators), "green" hydrogen can be produced [3, 8]. But it is important to realise that hydrogen is not a fuel source; it is an energy carrier, and has led to a worldwide development effort for hydrogen technology to power industrial, residential and transportation infrastructure – a concept known as the Hydrogen Economy [9].

On Earth hydrogen is rarely found in the pure form, but usually in a wide variety of inorganic and organic chemical compounds, the most common being water ($H_2O$). Hydrogen forms chemical compounds (hydrides) with nearly all other elements. Due to their ability to form long chains and complex molecules, combinations with carbon play a key role for organic life (hydrocarbons, carbohydrate) [10, 11].

The diatomic $H_2$ gas molecule can be produced from various sources. Currently, the most widespread process for the generation of hydrogen is the steam reforming from light carbohydrates, which, however, additionally generates undesirable $CO_2$ [12]. Alternatively, hydrogen can be obtained from water dissociation during electrolysis. This "green" hydrogen production from sustainable energy sources is a fully reversible process and it is a solution for sustainable ongoing hydrogen production for industries, storing "green" energy without increasing its carbon footprint and supplying energy for "green" mobility (transportation) [13].

In a Hydrogen Economy the lightest of all gases has to be processed like any other market commodity. It has to be packaged, transported by surface vehicles or pipelines, stored and transferred to the end user. There it can be converted back into electricity by fuel cells or other conversion devices. High-grade electricity from renewable or nuclear sources is needed not only to generate hydrogen by electrolysis but also for almost all essential marketing stages. At the end of the chain much less electricity becomes available than that has been invested upstream in the process of generation and marketing hydrogen [14, 15].

## 1.2 Hydrogen – Thermo-physical properties

Hydrogen is a chemical element with a chemical symbol H and atomic number 1. In normal conditions it is colourless, odourless and insipid gas, formed by diatomic molecules, $H_2$. It is the lightest element on the periodic table and the most abundant chemical substance in the universe. Hydrogen gas was first artificially produced in the early sixteenth century, via the mixing of metals with acids. English chemist Henry Cavendish was the first to identify the properties of hydrogen after he evolved hydrogen gas by reacting zinc metal with hydrochloric acid, in 1766 [16]. Seven years later, Antoine Lavoisier gave it the name "hydrogen," from the Greek words "ὕδωρ" and "γίγνομαι," which means "water-former" from its property to produce water when combusted [17].

Hydrogen consists of three isotopes, most of which is $^1H$. The ordinary isotope of hydrogen, *H*, is known as *protium*. In 1932, Urey announced the discovery of a stable isotope, deuterium ($^2H$ *or D*) with an atomic weight of 2. Deuterium is present in natural hydrogen to the extent of 0.015%. Two years later, an unstable isotope, tritium ($^3H$), with an atomic weight of 3 was discovered [18]. At normal conditions, hydrogen is in gaseous state. It has the second lowest boiling point and melting points of all substances, second only to helium. Hydrogen is a liquid below its boiling point of −253 °C (20 K) and a solid below its melting point of −259 °C (14 K) and atmospheric pressure. A phase diagram of hydrogen is shown in Figure 1.1. The boiling point of a fuel is a critical parameter since it defines the temperature to which it must be cooled in order to store and use it as a liquid [19].

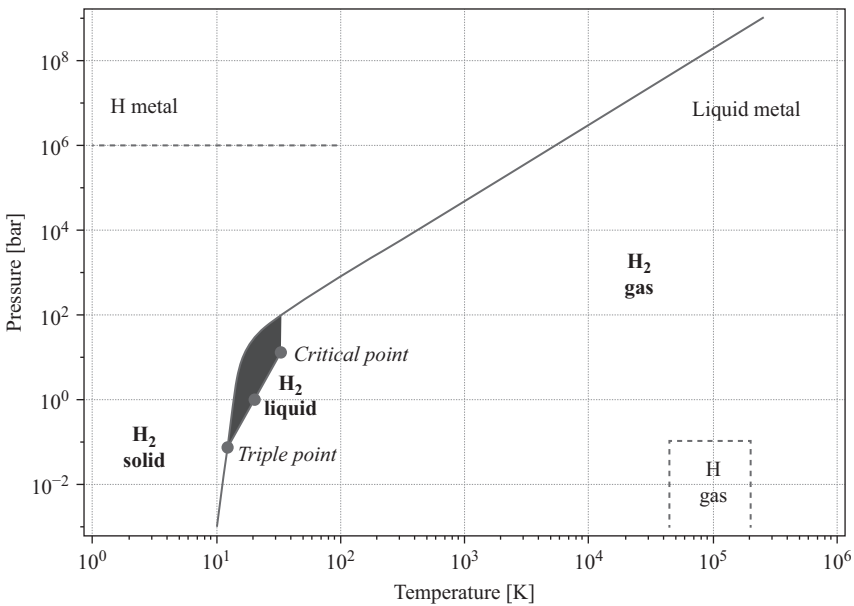

*Figure 1.1   Simple phase diagram for hydrogen [20]*

Table 1.1   Density of fuels in gaseous and liquid state [21]

| Substance | Gas at 20 °C, 1 bar (kg/m$^3$) | Liquid at boiling point, 1 bar (kg/m$^3$) |
| --- | --- | --- |
| Hydrogen | 0.09 | 70.8 |
| Methane | 0.65 | 422 |
| Gasoline | 4.4 | 700 |

Hydrogen has a molecular weight of 2.01594 g and a density of 0.08988 g/l at 0 °C and 1 bar, i.e. 7% of the density of the air. For comparison, the density of typical fuels is shown in Table 1.1. Density is measured as the amount of mass contained per unit volume. Density values only have meaning at a specified temperature and pressure since both of these parameters affect the compactness of the molecular arrangement, especially in a gas. At normal conditions, gaseous hydrogen is about 8 times less dense than methane, while in liquid state it is 6 times less dense than liquid methane and 55 times less dense than gasoline. Moreover, the volume ratio between hydrogen at 1 bar and compressed hydrogen at 700 bar is 440. Finally, it is interesting to note that more hydrogen is contained in a given volume of water or gasoline than in pure liquid hydrogen (111 kg/m$^3$, 84 kg/m$^3$ and 71 kg/m$^3$, respectively) [21].

The density of hydrogen at elevated pressure can be estimated using the principles of thermodynamics. While the behaviour of most gases can be approximated with a high accuracy by the simple equation of state of an ideal gas (PV = nRT), that relates the pressure, the volume and the temperature of a given substance, the behaviour of hydrogen deviates significantly from the predictions of the ideal gas model. The resulting deviation from the ideal gas law is always in the form of expansion – the gas occupies more space than the ideal gas law predicts. One of the simplest ways of correcting for this additional compression is through the addition of a compressibility factor, designated by the symbol Z. Compressibility factors are derived from data obtained through experimentation and depend on temperature, pressure and the nature of the gas. The Z factor is then used as a multiplier to adjust the ideal gas law to fit actual gas behaviour as follows [10]:

$$PV = nZRT \qquad (1.1)$$

By reducing the pressure P to the critical pressure $P_{cr}$ and the temperature T to the critical temperature $T_{cr}$, a generalised compressibility factor for all gases, can be drawn as a function of $P_R = P/P_{cr}$ and $T_R = T/T_{cr}$. The value of compressibility factor Z for hydrogen at high pressures and low temperatures in Figure 1.2 shows that, at ambient temperature, a value of 1.2 is reached at 300 bar, and at low temperatures even earlier. This means that a calculation of the hydrogen mass in a container from a measurement of temperature and pressure using the ideal gas equation will result in a mass 20% greater than in reality [10].

Energy is needed to compress gases and the compression work depends on the thermodynamically compression process, as well as on the nature of the gas. This is presented by the comparison of hydrogen with helium and methane in Figure 1.3.

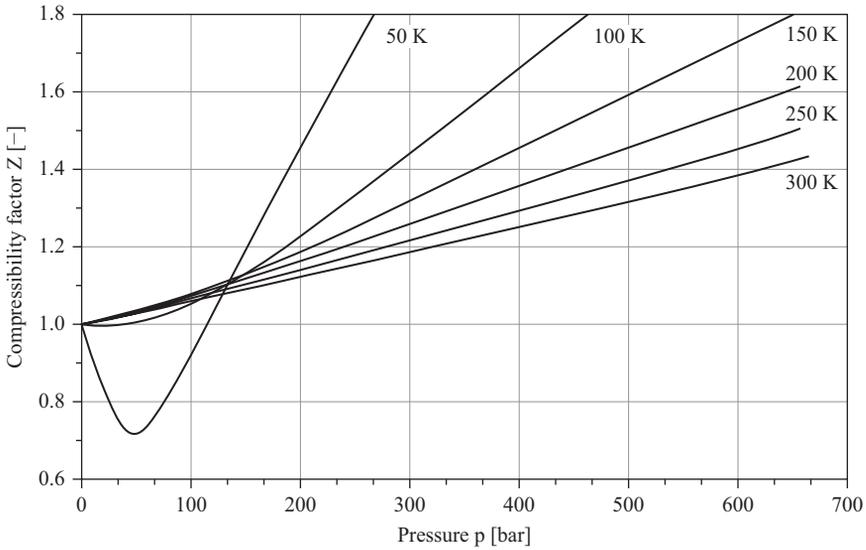

*Figure 1.2   Compressibility factor Z of hydrogen [10]*

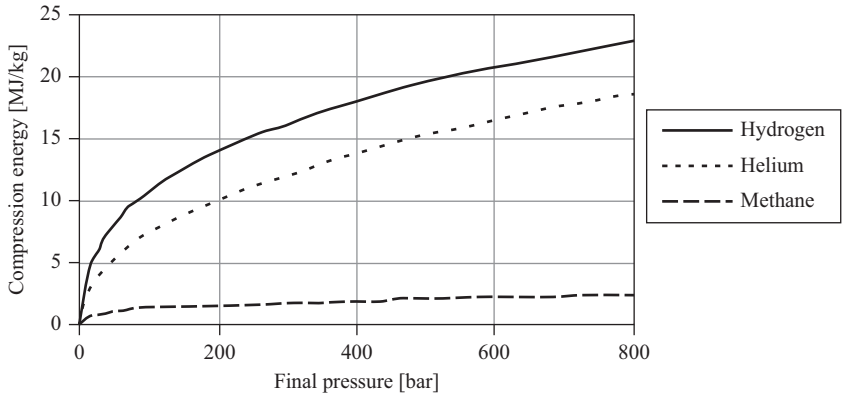

*Figure 1.3   Adiabatic compression work for hydrogen, helium and methane [22]*

The volumetric higher heating value (HHV; heat of formation) energy densities of different energy carrier options are illustrated in Figure 1.4. At any pressure, hydrogen gas clearly carries less energy per volume than methane (representing natural gas), methanol, propane or octane (representing gasoline). At 800 bar pressure, gaseous hydrogen reaches the volumetric density of liquid hydrogen. Therefore, compressed hydrogen cannot have the density of liquid hydrogen under any practicably achievable pressure conditions. But at any pressure, the volumetric energy density of methane gas exceeds that of hydrogen gas by a factor of 3.2. Furthermore, at 800 bar or in the liquid state, hydrogen must be contained in specially designed pressure tanks or cryogenic containers, while the liquid fuels are kept under atmospheric pressure conditions in simple containers [22].

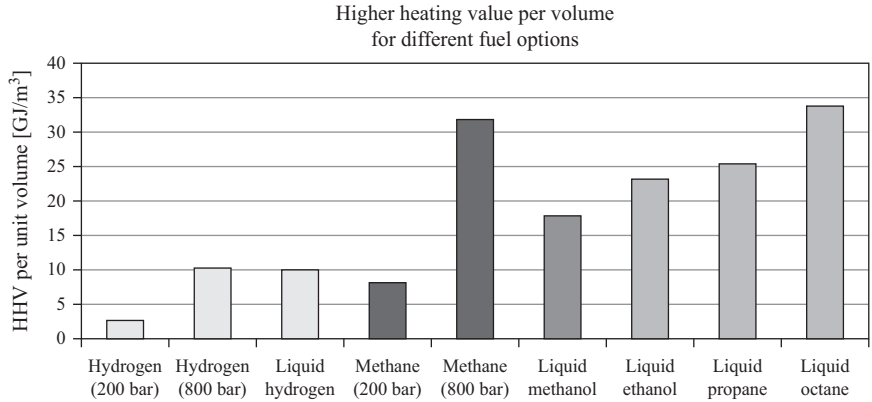

*Figure 1.4  Volumetric HHV energy density of different fuels [22]*

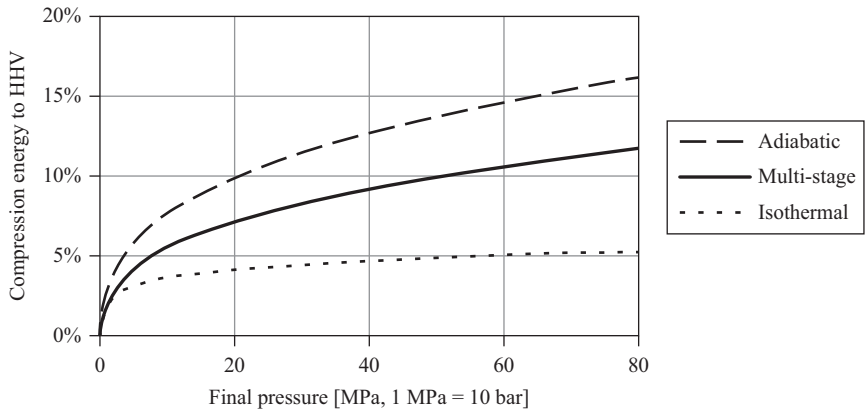

*Figure 1.5  Energy required for the compression of hydrogen compared to its higher heating value, HHV [22]*

Generally, the compressed gas has to be cooled down after each stage to make compression less adiabatic and more isothermal. Thus, hydrogen typically is compressed in several stages. The difference between adiabatic and isothermal ideal-gas compression of hydrogen is presented as in Figure 1.5. Multi-stage compressors with intercoolers operate between these two limiting curves. Also, hydrogen easily passes compression heat to cooler walls, thereby approaching isothermal conditions. It is worth to mention that, for a final pressure of 800 bar, the compression energy requirements would amount to about 13% of the energy content of hydrogen [22].

## 1.3 Hydrogen storage

Hydrogen can be stored using the following three basic storage methods [23]:

a. **Compressed hydrogen gas (CGH$_2$)**
b. **Liquid hydrogen (LH$_2$)**
c. **Solid storage of hydrogen (SSH$_2$)**

More specifically, since the first two methods can be applied by modifying hydrogen's physical state in gaseous or liquid form, the third method can be divided into the following categories:

- Physisorption in porous materials
- Absorbed on interstitial sites in a host metal
- Complex compounds
- Metals and complexes with water

### 1.3.1 Compressed hydrogen gas (CGH$_2$)

The most common storage system is high pressure gas steel cylinders, which are operated at a maximum pressure of 200 bar. But depending on the tensile strength of the cylinder material, higher pressures can be reached. New lightweight composite cylinders have been developed that are able to withstand pressures up to 800 bar, so that hydrogen can be reach a volumetric density of 36 kg/m$^3$, almost half as much in its liquid form at normal boiling point. The volumetric density of n-H$_2$ (normal hydrogen) as a function of the pressure at three different temperatures can be displayed as in Figure 1.6 [11]. It can be easily observed that hydrogen density does not follow a linear function over the increase of pressure. A hydrogen density of 20 kg/m$^3$ is reached at 300 bar. The volumetric density can be increased to around 40 up to 70 kg/m$^3$ by compressing the gas to a pressure of up to 700 or 2000 bar, respectively. However, 2000 bar is technically not feasible [25].

### 1.3.2 Liquid hydrogen (LH$_2$)

The second method to increase the volumetric density of hydrogen is, to decrease the temperature of the gas at a constant pressure to obtain the liquid phase. Liquid hydrogen is stored in cryogenic tanks at $-252\ °C$ (21 K) at ambient pressure. Because of the low critical temperature of hydrogen of $-240\ °C$ (33 K) (Figure 1.7), the liquid form can only be stored in open systems, as there is no liquid phase existent above the critical temperature, with simultaneous small quantity losses. The pressure in a closed storage system at room temperature could increase to $10^4$ bar. The simplest hydrogen liquefaction cycle is the Linde cycle, based on the Joule–Thomson effect [20].

In the Joule–Thomson effect, there is a coefficient, called Joule–Thomson coefficient, describes the extent and direction of the temperature change for an isenthalpic change of state (constant enthalpy h). The temperature that defines the transition from a positive to a negative Joule–Thomson coefficient (or the opposite) is called inversion temperature. A positive Joule–Thomson coefficient means that a decrease in temperature takes place along an isenthalpic pressure decrease, while a negative Joule–Thomson coefficient means that an increase in temperature takes place along an isenthalpic pressure decrease. Thus, the Joule–Thomson effect occurs when a gas or gas mixture experiences a change in temperature during an isenthalpic pressure change. It is worth to mention that an ideal gas does not show any Joule–Thomson effect.

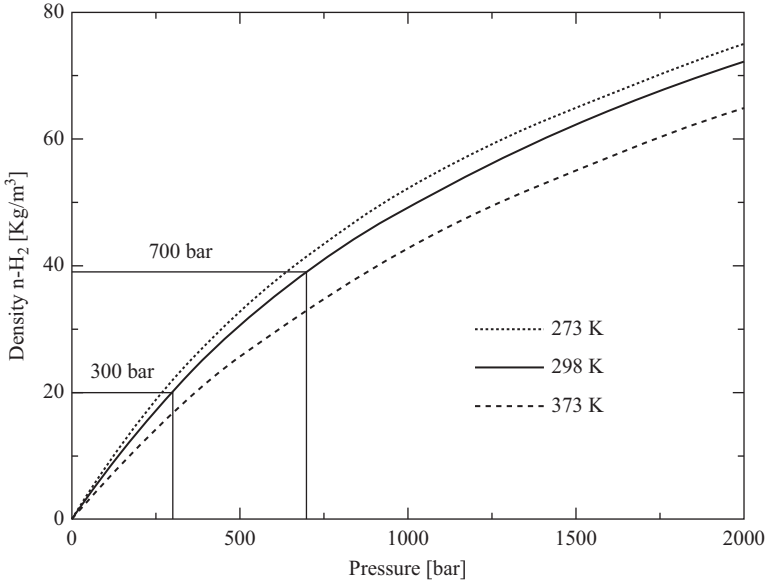

*Figure 1.6 Evolution of the volumetric density of n-H$_2$ as a function of pressure at three different temperatures [26]*

With ideal gases, the internal energy u and thus also the enthalpy h are only a function of temperature T. Thus ideal gases do not experience a change in temperature while the enthalpy remains constant, e.g. in a flow through a restriction. This means that the Joule–Thomson coefficient is zero. Hence, it can easily be judged from the T–s diagram (Figure 1.7) that, hydrogen can be regarded as ideal gas in an area where the isenthalpic lines (green lines) are horizontal.   AQ1

Hydrogen has a peculiar Joule–Thomson effect inversion temperature of around −83 °C (190 K) at a pressure below 50 bar. Consequently, a process other than a Joule–Thomson expansion has to be used to cool hydrogen in the range of 26.85 to −83.15 °C (300–190 K) [10]. For the liquefaction of hydrogen, the isenthalpic process of pressure decrease (throttling) is used. A throttling process proceeds along a constant-enthalpy curve (green line in Figure 1.7) in the direction of   AQ2
decreasing pressure, which means that the process occurs from left to right on a T–s diagram. If hydrogen is found in the region of liquid + gas phase after the throttling (region of horizontal isobaric lines in the T–s diagram), part of it becomes liquid. The whole process is like this: hydrogen is first compressed and then cooled (under constant pressure) in a heat exchanger, before it passes through a throttle valve where it undergoes an isenthalpic Joule–Thomson expansion, producing *some* liquid. The cooled gas is then separated from the liquid and returned to the compressor to undergo the same process.

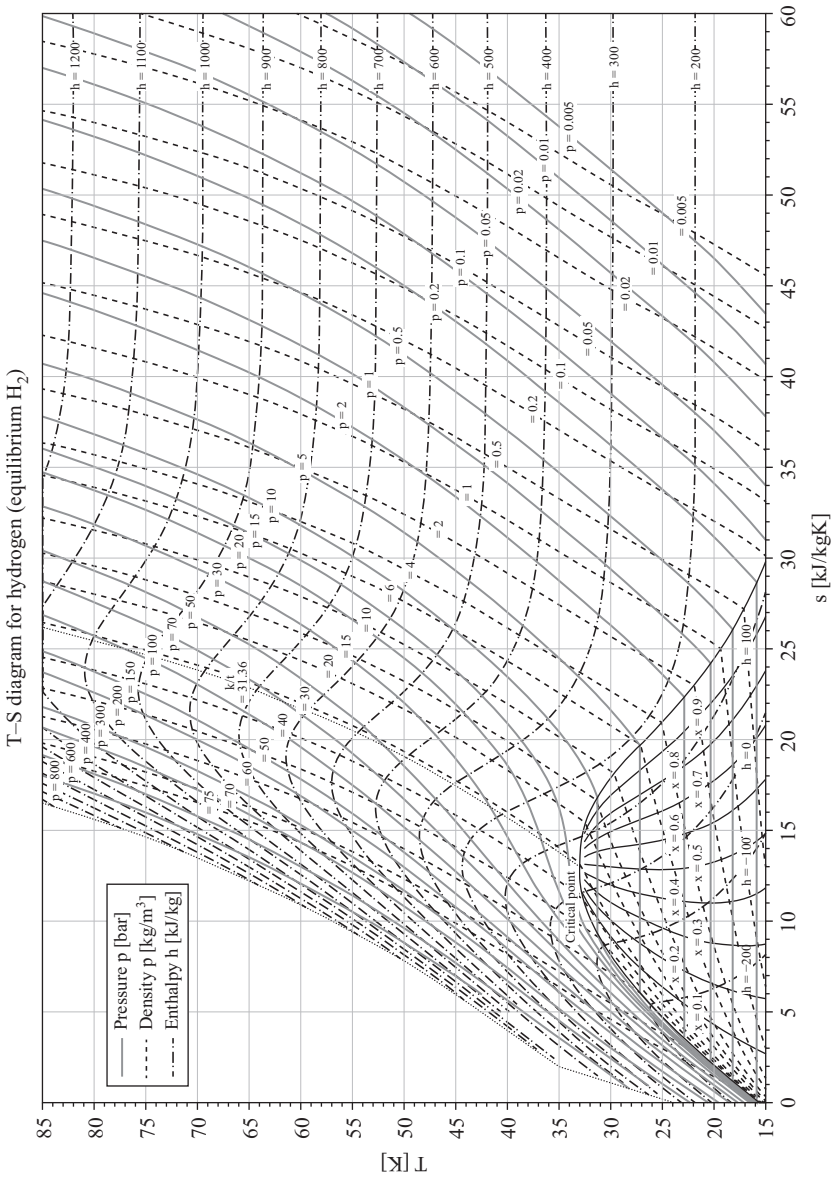

Figure 1.7  T–s diagram for equilibrium hydrogen for temperatures from −258 to −188 °C (15–85 K) [27]

In general, hydrogen liquefaction consumes large amounts of energy for achieving low temperatures and high pressures, as well as storage of liquid hydrogen for maintaining low temperatures at atmospheric pressure. The energy consumption, leakage along with safety issues led this solution to be applied only for special applications.

### 1.3.3  *Solid storage of hydrogen (SSH$_2$)*

As mentioned earlier, there are technologies that permit to store hydrogen directly by modifying its physical state into gaseous or liquid form. The traditional hydrogen storage facilities are complicated and moreover, liquid hydrogen requires the addition of a refrigeration unit to maintain a cryogenic state, thus adding weight and energy costs, and a resultant 40% loss in energy content [24, 28].

In the solid state storage there are various types of potential hydrogen storage materials. But as in every technological solution, there are always advantages and drawbacks, and the selection of a technology depends on the nature of the problem in a way that the disadvantages have less impact in the overall application.

Storage by absorption as chemical compounds or by adsorption on carbon materials has definite advantages from the safety perspective such that some form of conversion or energy input is required to release the hydrogen for use. A great effort has been made on new hydrogen storage systems, including metal, chemical or complex hydrides and carbon nanostructures [24].

Metal hydrides store atomic hydrogen in the bulk of the material. In the case of interstitial metal hydrides, the molecular hydrogen in the gas is split into atomic hydrogen on the surface of the material and then, it diffuses into the atomic structure of the host metal. In these hydrides, hydrogen acts as a metal and forms a metallic bond [29]. A large number of different metallic compounds exist that will absorb hydrogen in this manner. In most cases, however, the absorption does not occur at moderate temperature and hydrogen pressure for practical storage purposes and, the low mass of the absorbed hydrogen is only a small fraction of the mass of the host metal.

As hydrides are formed by direct reaction with gaseous hydrogen, it is important first to present the thermodynamics of hydride formation. The science and technology of reversible metal hydrides, in other words, the hydriding and dehydriding (H/D) of metals (M) by both direct dissociative chemisorption of H$_2$ gas (1) and electrochemical (2) splitting of H$_2$O are very simple. Many metals and alloys react reversibly with hydrogen to form metal hydrides according to the following reaction:

$$M + \frac{x}{2}H_2 \leftrightarrow MH_x + Q \tag{1.2}$$

$$M + \frac{x}{2}H_2O + \frac{x}{2}e^- \leftrightarrow MH_x + \frac{x}{2}OH^- \tag{1.3}$$

where M refers to a metal, a solid solution or an intermetallic compound, and Q to the heat of reaction. Since the entropy of the hydride is lowered compared to the metal and the gaseous hydrogen phase, the hydride formation is exothermic and the reverse reaction of hydrogen release is accordingly endothermic. Therefore, for hydrogen release/desorption heat supply is required.

There are several ways to show H-capacity. The reversible capacity, $\Delta(H/M)_r$, is conservatively defined as the plateau width, which can be considerably less than the maximum capacity, $(H/M)_{max}$. In practice, depending on available pressure and temperature ranges, engineering capacity is usually somewhere between $\Delta(H/M)$ and $(H/M)_{max}$. Capacity can be listed in either atomic H/M ratio or weight percent, both of which are used in some of the tables below. In calculating wt%, both H and M (i.e. not only M) are included in the denominator. In addition, it is sometimes useful to express capacity in volumetric terms, e.g. number of H atoms per unit volume (such as crystal cm$^3$). This measure is listed in some of the tables below as $\Delta NH/V$, where $\Delta NH$ represents the reversible capacity. Note that, this measure represents the volumetric density in crystal terms and does not include the void volumes inherent in engineering containers. In general, the mid-desorption plateau pressure, Pd, will be used in the graphs and tables below. Of course thermodynamics dictate the plateau pressures P must increase with temperature, usually close enough to the van't Hoff equation for engineering and comparison purposes,

$$\ln P = \frac{\Delta H^o}{RT} - \frac{\Delta S^o}{R} \qquad (1.4)$$

where, $\Delta H^o$ and $\Delta S^o$ are the enthalpy and entropy changes of the hydriding reaction (1), T is absolute temperature and R is the gas constant. For all of the hydrides to be discussed, $\Delta H^o$ and $\Delta S^o$ are negative, i.e. the hydriding (absorption) reaction is exothermic and the dehydriding (desorption) reaction is endothermic. The knowledge of $\Delta H^o$ especially is important to the heat management required for practical engineering devices and is a fundamental measure of the M–H bond strength.

The thermodynamic aspects of hydride formation from gaseous hydrogen are described by Pressure-Composition Isotherms (PCI), as presented in Figure 1.8. When solid solution and hydride phases coexist, there is a plateau in the isotherms, the length of which determines the amount of hydrogen stored. In the pure $\beta$-phase, the hydrogen pressure rises steeply with the concentration. The two-phase region ends in a critical point, a critical temperature noted as $T_c$, above which the transition from the $\alpha$- to $\beta$-phase is continuous.

The value of the enthalpy is an index of stability of a metal hydride. The higher the absolute value of the enthalpy shows a high degree of stability of the hydride, low dissociation pressure and the requirement of rather high temperatures to decompose it in order to liberate hydrogen [20].

At low concentration, as hydrogen initially diffuses into the pure metal interstices, the equilibrium pressure rises sharply, and a solid solution phase ($\alpha$-phase) is formed. The $\alpha$-phase has the same crystal structure as the bare metal. After the physical sites are saturated, a second phase, the metal hydride, begins to form.

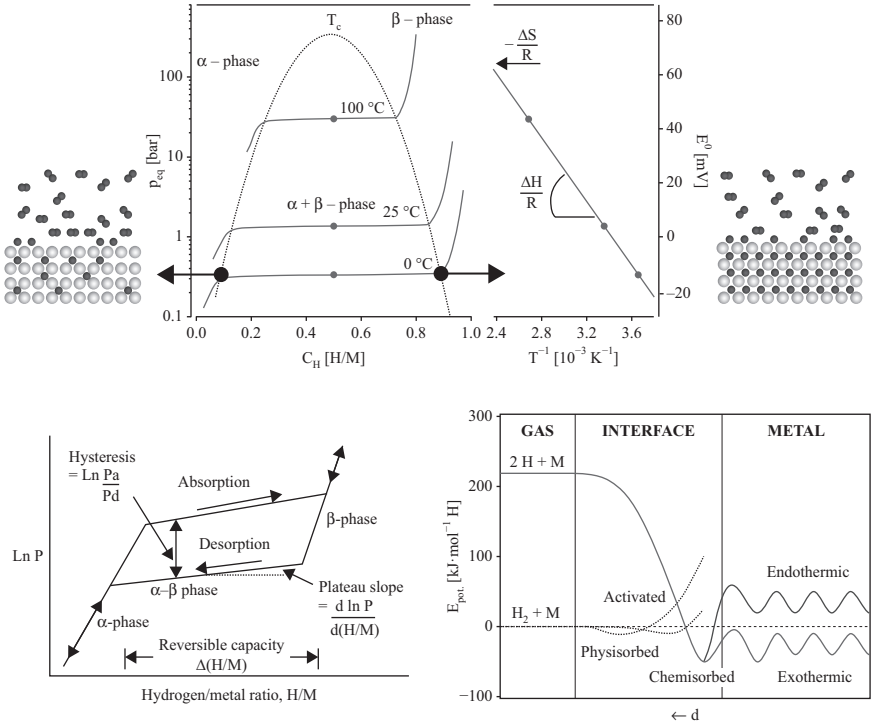

*Figure 1.8*   Schematic Pressure-Composition Isotherm, and van't Hoff plot giving the enthalpy of hydride formation (up). α is the solid solution of hydrogen and β is the hydride phase [20]. Full hydrogen absorption-desorption cycle (down-left) and Potential energies of $2H + M$ and $H_2 + M$ reactions at a gas-metal interface (down-right)

At this point, hydrogen is distributed methodically or randomly (depending on the hydride) in the metal host lattice and the concentration varies slowly with temperature. In this two-phase region the concentration increases while the hydrogen pressure is (ideally) constant [10]. Thus, a plateau is observed on the pressure versus hydrogen/metal ratio for fixed temperature. When the reaction is complete, another sharp pressure rise is observed if more hydrogen is added. At this point, nucleation of a higher concentration phase (β-phase) occurs. As the hydrogen pressure increases, the concentration also increases until the attractive H–H interaction becomes important [30]. The system now has three phases ($\alpha$, $\beta$ and hydrogen gas) and two components (metal and hydrogen). At a given temperature, the plateau pressure represents one point on a pressure-temperature (van't Hoff) Plot.

For practical purposes metal-hydrides are intermetallic compounds, which when exposed to hydrogen gas at certain temperatures and pressures, they absorb large quantities of hydrogen gas forming hydride compounds. The formed hydrides can then, under certain temperatures and pressures, desorb the stored hydrogen.

Hydrogen is absorbed interstitially in the metal lattice expanding the parent compound in the atomic and macroscopic level due to the Potential energies at a gas-metal interface. Such metal-hydrides can expand as much as 30% causing the decrepitation of the original ingots into fine powders. Metal-hydrides represent an exciting method of storing hydrogen. They are inherently safer than compressed gas or liquid hydrogen and have a higher volumetric hydrogen storage capacity. Some hydrides can actually store hydrogen in densities twice as much of that of the liquid hydrogen (0.07 g/cm$^3$).

There are a number of important hydride properties that must be considered in addition to the primary PCT properties. Some of the more important ones are listed below.

Activation is the procedure needed to hydride a metal the first time and bring it up to maximum H-capacity and hydriding/dehydriding kinetics. The ease of initial $H_2$ penetration depends on surface structures and barriers, such as the dissociation catalytic species and the oxide films. A second stage of activation involves internal cracking of metal particles to increase reaction specific surface. Decrepitation means the self-pulverisation of large metal ingots and large particles into powder (mostly single grain fine particles). This phenomenon results from a combination of hydriding volume change and the brittle nature of hydriding alloys. The morphology of the decrepitated powder affects heat transfer and also the tendency of powder migration into undesirable places in the hydrogen storage tank, like valve seats, assisting pipes. Unfortunately, most hydride powders have poor heat transfer coefficients and require engineering means for thermal enhancement (e.g. Al foam, internal fins). The morphology of the power can affect packing, which in turn can lead to internal gas impedance and container deformation.

Kinetics of hydriding and dehydriding can vary markedly from alloy to alloy. Fortunately, many room temperature hydrides have excellent intrinsic kinetics, so that the cycling of storage containers tends to be limited by heat transfer designs or accidental surface contamination. However, there are some materials that are kinetics limited, especially at low temperature.

Gaseous impurity resistance is a very important property, especially when the application is "open-ended" and uses new $H_2$ for each Hydriding/Dehydriding cycle, that $H_2$ often being impure. Depending on the alloy impurity combination, there can be several types of damage, such as Poisoning, Retardation, Reaction and Innocuous.

Usually, damages from poisoning and retardation are usually recoverable, but reaction damage is usually not. Cyclic stability is important and widely variable from alloy to alloy. Alloys and intermetallic compounds are usually metastable relative to disproportionation, the tendency to break up metallurgically (phase transformation) to form stable, not easily reversed hydrides. Even if very pure $H_2$ is used, disproportionation can occur with a resultant loss of reversible capacity. Safety usually centres on ignition, the tendency for a hydride powder to burn when suddenly exposed to air, e.g. an accidental tank rupture. But the term can also include toxicity resulting from accidental ingestion or inhalation. Alloy cost is influenced by several factors, including raw materials cost, melting and annealing costs, metallurgical complexities, profit and the degree of PCT precision needed for the particular application. It is important to point-out that the above factors in addition to raw materials cost can easily raise true alloy cost by more than 100%.

In addition to the $AB_5$, $AB_2$, $AB$ and $A_2B$ intermetallic compounds discussed above, several other families of intermetallics have been shown capable of reversible hydriding/dehydriding reactions. Examples include $AB_3$, $A_2B_7$, $A_6B_{23}$, $A_2B_{17}$, $A_3B$ and others. Most structures involve long-period $AB_5$ and $AB_2$ stacking sequences and are thus crystallographically related to these two classic families. Although none of these have attained commercial levels of interest, at least the $AB_3$ and $A_2B_7$ phases do have PCT properties, which are in the range of our interest. Most either have narrow plateaux with long sloping upper legs (e.g. $GdFe_3$) or multiple plateaux (e.g. $NdCo_3$ or $Pr_2Ni_7$). $La_2Mg_{17}$ was once reported to have 6 wt% H-capacity, recoverable at room temperature, but that claim has never been independently confirmed.

The solution of intermetallic compounds, which represent the hydrogen storage absorbed on interstitial sites in a host metal, will be further discussed. Research on intermetallic compounds for hydrogen storage has been commenced more than 30 years ago and it has always been on the foreground. The first discovery of hydrogen absorption by ZrNi [31], $LaNi_5$ [32, 30] and TiFe [33] opened new possibilities for industrial developments. However, for on-board storage, they remained at the stage of prototypes due to their weight penalty and low hydrogen storage capacity, since they don't meet the criteria determined by energy commissions around the world [34–39].

The host materials in the intermetallic compounds are ordered stoichiometric compounds, typically formed from two metallic components, A and B. As it is illustrated in Figure 1.9, intermetallic compounds are built by alloying a metal which easily forms stable hydrides (A) and another element which does not form stable hydrides (B). The components A and B can generally be fully or partially substituted by other elements of relatively similar size or chemistry. The intermetallics thus formed could then be grouped according to their stoichiometry such as $AB_5$, $AB_2$, $AB$ and others. Hydrogen absorbing intermetallics form a number of different groups as shown in Table 1.2.

In general, metals are crystallised in one of the following structures in near ambient conditions (Figure 1.10):

- Body Centred Cubic (BCC)
- Face Centred Cubic (FCC)
- Hexagonal Closed Packed (HCP)

More specifically, many intermetallic compounds of composition AB mainly crystallise in the BCC structure, while $AB_2$ crystallise in one of three closely related structures (Laves Phases), which are:

- The hexagonal C14 structure, typified by the phase $MgZn_2$
- The cubic C15 structure, typified by the phase $MgCu_2$
- The hexagonal C36 structure, typified by the phase $MgNi_2$

These components, the "Laves phases," involve the association of large A atoms and small B atoms. Since the ratio of the atomic diameters of the components $d_A/d_B$ is approximately 1.2, the phases are regarded as controlled essentially by atomic size relationships. This interpretation is supported by the observation that the A and B components may be taken from any group of the Periodic Table, while the same element may form either A or the B component in different compounds [42].

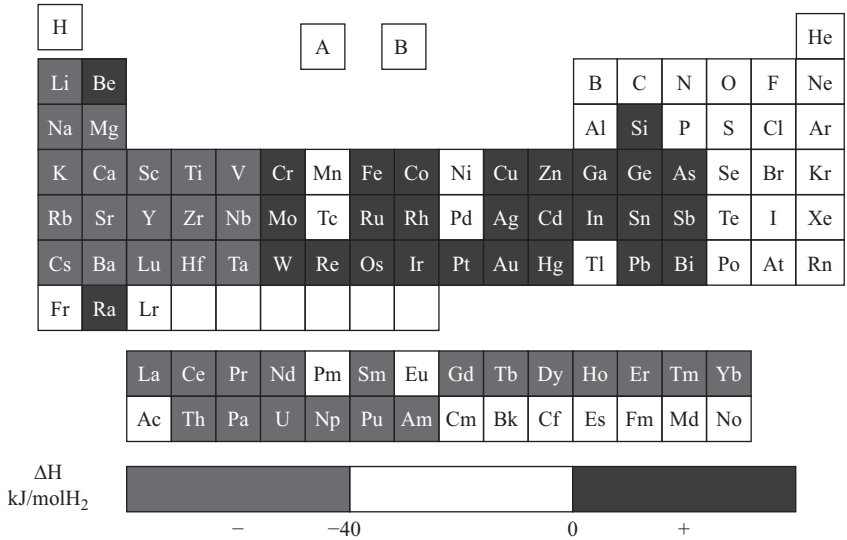

Figure 1.9  Hydride and non-hydride forming elements in the periodic system of elements [40]

Table 1.2  Examples of intermetallic hydrides [20, 40]

| Composition | A | B | Compounds |
|---|---|---|---|
| $AB_5$ | Ca, La, rare earth | Ni, Cu, Co, Pt, Fe | $CaNi_5$, $LaNi_5$, $CeNi_5$, $LaCu_5$, $LaPt_5$, $LaFe_5$ |
| $AB_2$ | Zr, Ti, Y, La | V, Cr, Mn, Fe, Ni | $LaNi_2$, $YNi_2$, $YMn_2$, $ZrCr_2$, $ZrMn_2$, $ZrV_2$, $TiMn_2$ |
| $AB_3$ | La, Y, Mg | Ni, Co | $LaCo_3$, $YNi_3$, $LaMg_2Ni_9$ |
| $A_2B_7$ | Y, Th | Ni, Fe | $Y_2Ni_7$, $Th_2Fe_7$ |
| $A_6B_{23}$ | Y | Fe | $Y_6Fe_{23}$ |
| AB | Ti, Zr | Ni, Fe | TiNi, TiFe, ZrNi |
| $A_2B$ | Mg, Zr | Ni, Fe, Co | $Mg_2Ni$, $Mg_2Co$, $Zr_2Fe$ |

During hydrogenation, hydrogen atoms will occupy specific interstitial sites. The interstitial sites in the three major crystal structures of hydrides are shown in Figure 1.10. Only octahedral (O) and tetrahedral (T) sites are shown because they are the only ones occupied by hydrogen atoms. Nevertheless, some distinction should be made between the different crystal structures. In the FCC lattice, the T and O sites are enclosed in regular tetrahedral and octahedral, respectively, formed by metal atoms. In the HCP lattice, the tetrahedral or octahedral sites become distorted as the ratio of lattice parameters c/a deviates from the ideal value of 1.633 [10].

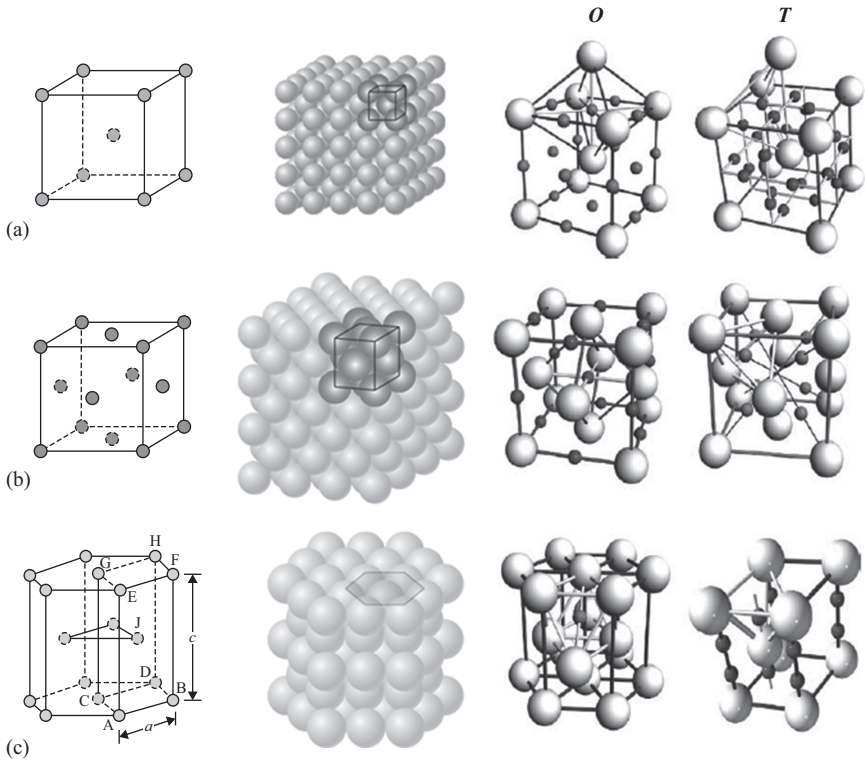

Figure 1.10   Schematic representation for the (a) Body Centred Cubic (BCC), (b) Face Centred Cubic (FCC) and (c) Hexagonal Closed Packed (HCP) structure, and their interstitial octahedral (O) and tetrahedral (T) sites [10, 41]

An important feature of the $AB_2$ alloys (and to a lower extent, the AB alloys) is that their operating characteristics (equilibrium pressure and temperature) can be adjusted through solute substitution for both A- and B-site metals. The equilibrium pressures at a selected temperature can often be changed by more than two orders of magnitude. In general, the larger the lattice parameters of the host alloy, the lower the pressure. Deviations in the alloy composition (i.e. changes in the B/A ratios) also lead to significant changes in the PCI characteristics, which can be advantageous in many situations. Unfortunately, both of these alloy variations usually result in a smaller hydrogen storage capacity than pure stoichiometric binary $AB_2$ intermetallics.

### 1.3.3.1   AB-type TiFe intermetallic compounds

Among hydrogen storage materials, TiFe has the advantages of a large absorption capacity (1.8–1.9 wt%), good kinetics for sorption and desorption of hydrogen after activation, suitable equilibrium pressures (5–10 bar) and low cost. However, TiFe has the great disadvantage of not being easily activated at room temperature [33], [43–47]. But despite the advantages of TiFe, its weak points must be overwhelmed in order for this compound to find a practical application. As it is previously mentioned, a common method to modify the hydrogen sorption properties of a hydride is the insertion of substituting elements. For this compound, V, Mn or/and other transition metals are widely used to partially substitute Fe. For example, when a small amount of Mn is substituted for Fe in the TiFe parent compound, activation is promoted and the resistance to contamination by impure gases is increased. TiFe intermetallic compound counts down a history of about 40 years since its discovery as a potential hydrogen storage material and has been thoroughly studied ever since in an effort to improve its properties [48–52].

### 1.3.3.2   $AB_2$ Laves Phases Zr-based intermetallic compounds

Laves phases form the largest group of intermetallics and thus have a wide range of properties. Their stability depends on various factors such as: geometry, packing density, valence electron concentration or the difference in electronegativity. Laves phases have been recognised to be attractive hydrogen storage materials, particularly the Zr-based alloys. They have relatively good hydrogen storage capacity and kinetics, long cycling life and low cost but they are usually too stable at room temperature and are sensitive to gas impurities. Shaltiel *et al.* [54] first studied the hydrogenation characteristics of various Zr-based Laves Phase compounds such as $ZrV_2$, $ZrCr_2$, $ZrMn_2$, $ZrFe_2$ and $ZrCo_2$. Most of these alloys could absorb large quantities of hydrogen but the hydrides formed were too stable to be of practical significance. Therefore many studies performed so far have paid attention to increasing the equilibrium pressure of Zr-based Laves Phase compounds, without modifying the absorption capacity, by partial substitution of A or B element by other elements. Substitution usually involves transition elements, creating multi-component systems ($Zr_{1-x}X_x$(Fe, Mn, Cr)$_{2-y}Y_y$, where X = Ti, Y, Hf, Sc, Nb and Y = V, Mo, Mn, Cr, Fe, Co). However, in many cases Ti and Fe are being mostly used for the substitution of A and B elements, respectively [10], [53–59].

## 1.4   Hydrogen compressors

Hydrogen compression is achieved by a mechanical device that increases the pressure of the gas by reducing its volume. However, the low molar weight of hydrogen requires the use of a volumetric compressor rather than a centrifugal one in order to gain efficiency. Moreover, the energy used to compress gas does not only produce a pressure increase, but also generates heat. Hydrogen compressors are categorised in mechanical, non-mechanical hydrogen compressors. Their properties and characteristics will be further described.

### 1.4.1 Mechanical compressors

The piston compressor is an electro-hydraulically driven, non-lubricated, liquid-cooled, single-stage unit, including an electric motor, hydraulic oil tank, high pressure gas intensifier and intensifier shifting mechanism. As it can be seen in Figure 1.11, representing a single-stage unit, the intensifier contains a hydraulic drive cylinder in the centre that is coupled by tie rods with two single-stage gas cylinders on each side. The fluid power drive provides the intensifier with high pressure hydraulic oil. During operation, the gas fills the cylinder and then, the force of the hydraulic pressure acts on the hydraulic piston compressing the gas in the cylinder. Once compression is completed, the four-way valve redirects the hydraulic fluid and the piston assembly moves in the opposite direction.

In the case of a piston-metal diaphragm hydrogen compressor, the gas from the piston and related components is isolated by a set of metal diaphragms as shown in Figure 1.12. The piston moves a column of hydraulic fluid, which in turn moves the diaphragm set and displaces the gas to be compressed. As a consequence, the process is more isentropic than adiabatic, thus allowing to achieve a higher compression ratio [11].

### 1.4.2 Non-mechanical compressors

These types of compressors have several advantages over mechanical hydrogen compressors, including smaller size, lower capital, operating and maintenance costs and the absence of moving parts, which eliminates problems related to wear, noise and intensity of energy usage. Since hydrogen is completely separated from the hydraulic fluid, high purity hydrogen can be supplied [11].

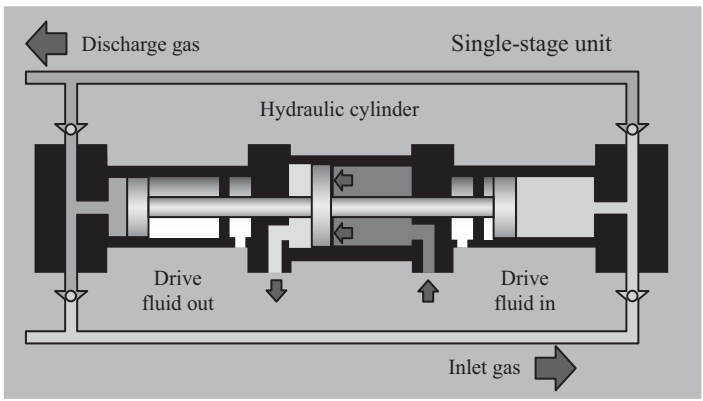

Figure 1.11  Schematic diagram of a single-stage compression unit [11]

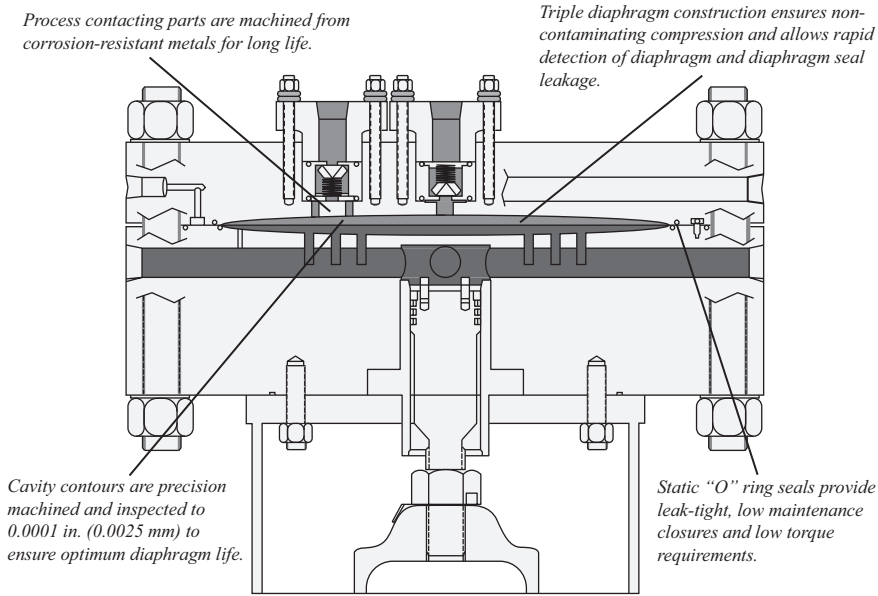

*Figure 1.12 Schematic diagram of a metal diaphragm hydrogen compressor [47]*

As mentioned earlier, this category of compressors is represented by two types of devices, the electrochemical and the metal hydride hydrogen compressor [60, 61]. Within the aim of this study, the electrochemical compressor will be briefly presented while the metal hydride hydrogen compressors will be thoroughly discussed.

The **electrochemical hydrogen compressor** is used when small quantities of hydrogen have to be delivered in high pressure, because it is more efficient than the mechanical compressor in this regime [42]. The working principle is based on an electrochemical cell, composed of an anode, a membrane electrode assembly and a cathode, as it schematically represented in Figure 1.13. When a potential difference is applied, the hydrogen at a pressure Pa is oxidised to H+ at the anode. These ions are transported through the membrane to the cathode, where they are reduced to hydrogen at a pressure Pc > Pa, if the cathode compartment is hermetically sealed. As long as hydrogen and power are supplied, this electrochemical reaction continues to compress hydrogen. A multi-stage electrochemical hydrogen compressor incorporates a series of membrane electrode assemblies. It should be noted that the process is selective for hydrogen, as the inert gas components cannot pass the membrane [11].

On the other hand, **metal hydride hydrogen compressors** are efficient and reliable thermally powered systems that use the properties of reversible metal hydride alloys to compress hydrogen without contamination. The most important component of this thermal engine is the metal hydride. A wide range of different operation characteristics can be obtained by either varying the type of the hydride or just modifying the formula of a reference alloy. The operating principle of the metal hydride hydrogen compressor is based on heat and mass transfer (hydrogen) in the reaction bed during absorption and desorption process [11]. The advantages of metal hydride compression include simplicity in design and operation, absence of moving parts, compactness, safety and reliability, and on the most important, the possibility to consume waste industrial heat instead of electricity [63].

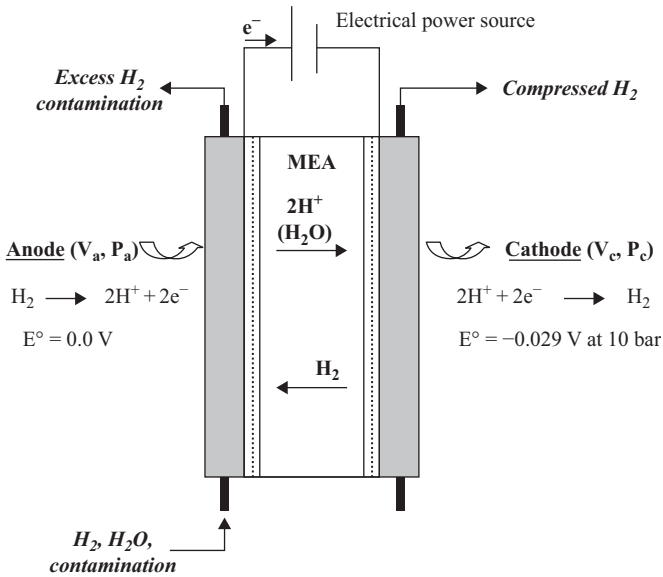

Figure 1.13  Principle of the electrochemical hydrogen compressor [62]

The first studies on metal hydride hydrogen compressors have been reported already from 1984, and since then numerous R&D activities on the development of such a device have been published in several forms, as papers, conference presentations, reports and patents [63–83].

Since this type of compressor is based on metal hydrides, as a consequent, it follows the rules of function of hydrides. That means that they have in common all the advantages and drawbacks. But more specifically, metal hydride compressors are simple and efficient pressure/temperature swing absorption-desorption systems. This allows not only controlling pressure by varying the temperature, but it also opens a new horizon for hydrogen separation and purification [63].

The selection of an adapted metal hydride alloy is important to reach the desired performance of the compression cycle. The thermodynamic ($\Delta H$ and $\Delta S$), thermal (specific heat $C_p$) and kinetic properties of the selected material should be characterised in detail in order to optimise compressor operation.

A single-stage thermal compressor is composed of a module filled with a reversible metal hydride alloy. Figure 1.14 displays the operation of the hydride compressor, which comprises four processes [11]:

Point **A**: Hydrogen is absorbed into the alloy bed at low temperature ($T_c$) and at low pressure ($P_s$).

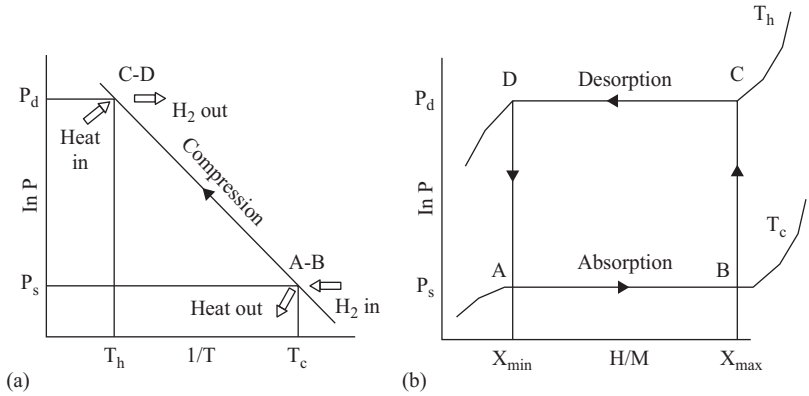

*Figure 1.14 Principle of operation of a single stage metal hydride hydrogen compressor [75]*

- Point **B**: The module is subsequently heated from ($T_c$) to a higher temperature ($T_h$) with compression taking place at the same time.
- Point **C**: Compressed hydrogen is released at high temperature ($T_h$) and high pressure ($P_d$).
- Point **D**: The module is cooled down (from $T_h$ to $T_c$).

In a multi-stage metal hydride compressor, hydrogen is successively absorbed and desorbed into and out of several hydride beds. The hydrides operate between two temperature levels. The warm side provides the heat for desorption of hydrogen, and the cold side is the heat sink to dissipate the heat generated when hydrogen is absorbed. The hydride alloys are selected in such a way that the lower pressure hydride desorbs hydrogen at high temperature into the higher pressure hydride which absorbs hydrogen at low temperature (Figure 1.15) [84].

But the selection of the alloys should take into account more factors, mainly concerning the operation conditions of the device. Thus, before the selection it is essential to set the boundary conditions, such as:

- **Cold side temperature** (or temperature of cooling). The value of this parameter is usually around ambient temperature or less, depending on the procedure and the resources available for the cooling.
- **Hot side temperature** (or temperature of heating). Since water is mainly used for heating and cooling as well, the value of this parameter is not usually exceeding 90 °C, but if another way of cooling and heating is installed, this temperature can be as high as the metal hydride thermodynamic properties allow. On the other hand, there is an excess of waste heat from other processes that is available, but it hardly surpasses 70–80 °C (waste heat of an electrolyser and fuel cell/solar collector).
- **Input pressure**. This parameter is independent only if the metal hydride hydrogen compressor is not coupled with another device. But in order to increase the overall efficiency of the device, the aim is to couple such a compressor with an electrolyser; the input pressure must be in the range of the output pressure of the electrolyser, i.e. 10–30 bar.
- **Output pressure**. The value of this parameter is dependent of the technical specifications of the device, which includes the hot side temperature and the capacity of the hydride beds to withstand the temperature and the pressure demanded.

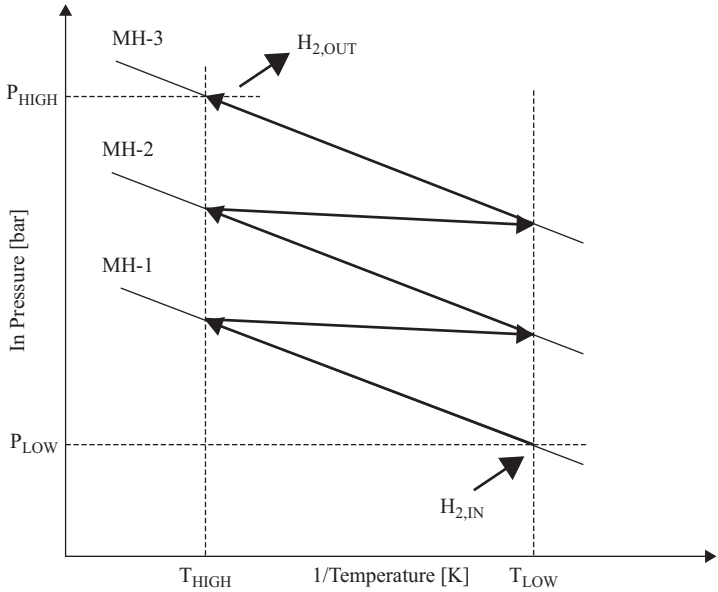

*Figure 1.15   Principle of a three-stage metal hydride hydrogen compressor [71]*

When the discussion comes to the overall hydrogen compression by metal hydrides unit, more factors must be taken into account. The major factor affecting hydrogen compression efficiency is the plateau slope of the materials selected. In multi-component intermetallic compounds the sloping plateaux are originated form compositional fluctuations due to the presence of impurities or because of fluctuations of the stoichiometry ($AB_{n\pm x}$) within the homogeneity region [85]. In addition to operating pressure-temperature ranges, an important parameter of the material for hydrogen compression is the process productivity. Another important factor for the hydrogen compression performance of the real metal hydride systems is hysteresis, as the values of plateau pressures for hydrogen absorption/hydrogenation are higher than the ones for hydrogen desorption/dehydrogenation. Hysteresis is caused by various reasons; the most frequent one is the stresses which appear in the course of growth of hydride nuclei inside the matrix of the metal alloy having lower molar volume [63].

Many groups have been focused in hydrogen compression metal hydrides materials systems as it is recognised as one of the potential efficient ways of hydrogen storage technology in mobile applications [86–90].

## Acknowledgements

I would like to thank my ex-PhD students, Dr Alexandra Ioannidou, Dr Evangelos Gkanas and Dr Evangelos Koultoukis, for their collaboration over the last decade in metal hydride storage and compression experimental and theoretical research work.